# IMPACTO DO CÂNCER NA ATIVIDADE CEREBRAL

THE IMPACT OF CANCER ON THE NEURAL ACTIVITY

Kelly Cristiane Iarosz[1*], Antonio Marcos Batista[2], Murilo da Silva Baptista[3], Paulo Ricardo Protachevicz[4]

[1]University of Aberdeen, Institute for Complex Systems and Mathematical Biology (ICSMB), Pós-doutoranda CNPq – Brasil – <u>kiarosz@gmail.com</u>

[2]Universidade Estadual de Ponta Grossa, Departamento de Matemática e Estatística – antoniomarcosbatista@gmail.com

[3]University of Aberdeen, Institute for Complex Systems and Mathematical Biology (ICSMB) – murilo.baptista@abdn.ac.uk

[4]Universidade Estadual de Ponta Grossa, Programa de Pós-Graduação em Ciências/Física, Pós-graduando (Mestrado) – protachevicz@gmail.com

## RESUMO

Apresenta-se neste trabalho, um estudo da influência da perda de neurônios na taxa de disparos neuronais. Para tal estudo foi elaborado um modelo de autômato celular que simula a proliferação de células cancerígenas em um tecido cerebral e a morte de neurônios devido a falta de assistência das células que lhes dão suporte nutricional e funcional. Por meio do modelo analisou-se a taxa de disparos neuronais considerando diferentes valores de perturbação externa e probabilidades de proliferação de células cancerígenas. Com este trabalho conclui-se que a presença de proliferação não controlada de células, diminui a taxa de disparos neuronais.

 **Palavras chave:** neurônio; células glias; disparos.

## ABSTRACT

We study the impact of the decrease in the neural population on the neuronal firing rate. We propose a cellular automaton model from cancerous growth in a brain tissue and the death of neurons due absence of cells that help support to neurons. We use this model to study how the firing rate changes when the neuronal networks is under different external stimuli and the cancerous cells have different proliferation rate. Our work shows that the cancer proliferation decreases the neuronal firing rate.

# 1. Introdução

A modelagem matemática de sistemas biológicos é uma área em expansão, sendo um dos campos mais promissores da ciência para os próximos dez anos [Cookson2011]. Diversas ferramentas podem ser utilizadas para modelagem matemática, para este trabalho escolheu-se o autômato celular.

Os autômatos celulares são modelos considerados simples por apresentar apenas regras de interação e respostas da vizinhança [Wolfram1984, Wolfram1985]. As simulações de interações celulares feitas com autômatos são bem vistas, por se mostrarem diretas, ter baixo tempo computacional e serem possíveis mesmo sem o uso de super máquinas ou processamento de última geração. Devido a esse conjunto de características, os autômatos vêem sendo utilizados em simulações dos mais diversos sistemas, como exemplo pode-se citar a descrição de tráfego [Gao2007], proliferação celular [Iarosz2010, Iarosz2011], ecossistemas [Manor2008], criticalidade auto-organizada em plasma [Santos-Lima2012], arquitetura [Borries2007], redes neurais [Copelli2002, Iarosz2012], etc.

Sendo assim, propõem-se mostrar um estudo sobre o comportamento da taxa de disparos neuronais quando um tecido cerebral tem a presença de proliferação não controlada de células. As células em questão são chamadas de gliais, tais células funcionam de modo integrado com os neurônios [Fieldes2006]. Enquanto as células do tipo gliais tem funções específicas como fornecer nutrientes e oxigênio, isolar um neurônio do outro e participar nas transmissões sinápticas, regulando a liberação de neurotransmissores ou moderando funções [Rajkowska2000, Laming2000, Griffith1996] os neurônios são altamente especializados na transmissão de informações na forma de impulsos nervosos [Lent2010]. Como ambos são integrados, caso o conjunto de células que fornece a base para sobrevivência dos neurônios venha sofrer alguma alteração, consequentemente os neurônios também sofreram [Kandel2000].

O trabalho está fundamentando em pesquisas anteriores, nas quais Iarosz e colaboradores apresentam uma série de estudos sobre proliferação de células cancerígenas, evolução temporal do número de células cancerígenas, supressão de células cancerígenas, comportamento temporal das células cancerígenas em função do tamanho de rede, tempo para que a metástase venha a ocorrer [Iarosz2010, 2011], efeitos da probabilidade de conexão não-local nas taxas médias de disparos dos neurônios acoplados, auto-sustentabilidade, valor de probabilidade crítica das conexões em função do tamanho da rede e fator de amplificação [Iarosz2012].

As laudas estão organizadas da seguinte forma: a seção 2 refere-se ao modelo de autômato celular e as regras utilizadas, na seção 3 apresenta-se uma explicação sobre as células neuronais, a proliferação celular e os disparos neuronais, seção 4 reporta-se aos parâmetros utilizados nas simulações, na seção 5 apresentam-se os resultados do modelo proposto e finalmente na seção 6 encontram-se as conclusões.

**2. Autômatos Celulares**

Os autômatos celulares são modelos matemáticos discretos no tempo, no espaço e nas variáveis dinâmicas [Wolfram1984, Wolfram1985]. Com a evolução de um autômato celular observa-se a interação entre as vizinhanças e com isso surgem comportamentos complexos [Wolfram2002] com estados variando conforme regras atribuídas ao modelo [Wolfram1983a, Wolfram1983b].

Na Figura 1(a) representa-se três posições, sendo uma posição central ($a_0$), sua vizinhança a esquerda ($a_{-1}$) e a direita ($a_1$), conforme são atribuídas regras de interação entre as três posições, um autômato celular unidimensional vai mostrando sua evolução. Para cada posição são atribuídas características denominadas estados. O mais comum em relação aos estados é chamar de ocupado ou não ocupado, mas isso não é uma regra geral, e tudo depende do que está sendo modelado, como exemplo, na Figura 1(b) tem-se uma posição ocupada, referente a $a_0$ da Figura 1(a) e duas vazias, referentes a $a_{-1}$ e $a_1$. Estas evoluem da seguinte forma: se a posição $a_0$ encontra-se ocupada no primeiro instante de tempo, então, no segundo instante de tempo é ficará vazia e sua vizinha diretamente a direita passará a ser ocupada, como mostra a Figura 1(b).

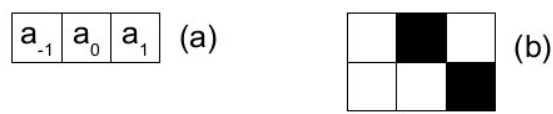

**Figura 1** – Formação de um autômato celular com $a_0$ representando a posição central e as vizinhas $a_{-1}$ e $a_1$ imediatamente a esquerda e a direita em (a) e em (b) observa-se uma possível evolução segundo regras atribuídas.

O autômato celular criado para o presente trabalho tem regras para a proliferação de células, e para o estudo de taxa de disparos neuronais, sendo assim, tem suas posições ocupadas por células neuronais e os estados de cada célula variam de acordo com suas funções.

*2.1 Regras para Proliferação de células*

A condição lógica de proliferação celular ocorre quando de uma célula mãe, originam-se duas outras células com igual material genético, chamadas filhas [Maton1997]. Porém, quando algum distúrbio ocorre no mecanismo que controla a duplicação, as células deixam de apresentar uma proliferação regulada e passam a se duplicar de forma incontrolada [Bruce2010]. Na maioria dos casos essa falta de regularidade é causado pelo acúmulo gradual de mudanças genéticas, causando neoplasias, comumente chamada de tumor [Hahn2002, Herren1968].

Na Figura 2 observa-se a representação das regras utilizadas neste trabalho para transição dos estados das células. Basicamente, as células podem se duplicar ou sofrer uma transformação de acordo com probabilidades. Na Figura 2(a) Uma célula tem a probabilidade $p_1$ de duplicação, como a probabilidade utilizada como parâmetro não se encontra na faixa de proliferação normal de uma célula, então, ela será uma célula cancerígena $C$. Na Figura 2(b), mostra-se que tanto células cancerígenas $C$ quanto células citotóxicas que fazem parte da reação natural do organismo $C_1$, tem uma probabilidade $p_2$ de transformação em células com funções celulares suprimidas $C_2$. As células $C_2$, por sua vez tem probabilidade $p_3$ de retornar a $C_1$ e/ou transformarem-se em células mortas $M$. Em (c) observa-se a probabilidade $p_3$ de dissolução das células mortas.

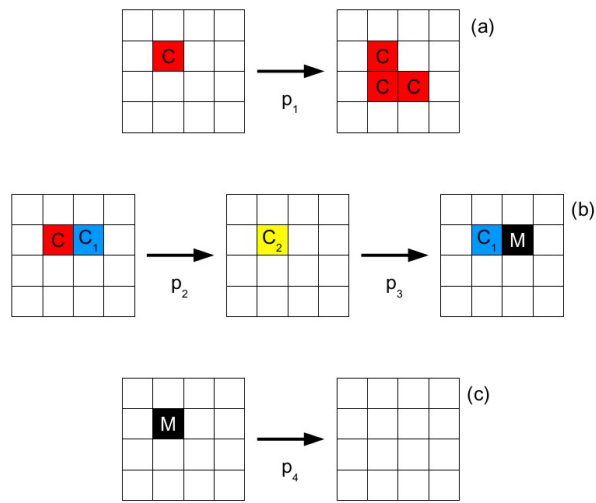

**Figura 2** – Autômato celular com regras de proliferação para células cancerígenas $C$, células citotóxicas $C_1$, células resultantes do processo citotóxico $C_2$ e células mortas $M$. Em (a) tem-se a probabilidade $p_1$ para $C \Rightarrow 2C$, em (b) encontra-se a probabilidade $p_2$ para $C + C_1 \Rightarrow C_2$ e a probabilidade $p_3$ para $C_2 \Rightarrow C_1 + M$, em (c) tem-se a probabilidade $p_4$ de dissolução das células mortas $M \Rightarrow$ .

*2.3 Regras para taxa de disparos neuronais*

Os neurônios são células altamente especializadas em transmissão de informações na forma de impulsos nervosos [Stell1977]. Os impulsos nervosos são resultado de alterações químicas nas superfícies da membrana plasmática [Chialvo2004] e passa de neurônio para neurônios com ajuda de mediadores até chegar no sistema nervoso central [Gerstner2002]. O ponto de união entre os neurônios denomina-se sinapse, e as mesmas podem ser de natureza elétrica ou química.

Na construção das regras para a atividade neuronal no modelo, as posições encontram-se em um estado ativo, representando o um neurônio disparando, ou em um estado inativo representando um período refratário, sem atividade neuronal [Erichsen1993]. A ligação com a vizinhança representa o tipo de conexão entre os neurônios, a interação com os vizinhos mais próximos representam as sinapses elétricas, são conexões locais (curtas) e bidirecionais. A interação com os vizinhos mais distantes é feita por meio de conexões não-locais, representando as sinapses químicas, que por sua vez são longas e unidirecionais. A proposta das conexões não-locais é baseada no conceito de atalhos proposto por Newman e Watts [Newnan1999a, 1999b], enquanto que as locais está diretamente ligada com os vizinhos mais próximos. Um neurônio pode sofrer influência devido ao estímulo externo quando um ou mais vizinhos da i-ésima célula neuronal estiverem disparando e a perturbação externa ao i-ésimo neurônio for diferente de zero [Copelli2002]. Essa lógica de disparos neuronais é apresentada na Figura 3.

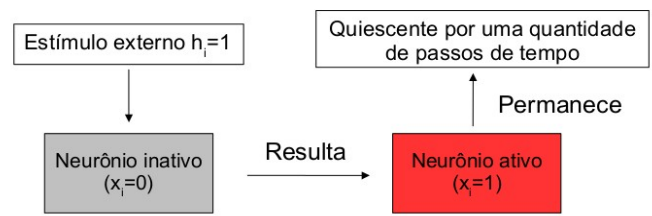

**Figura 3** – Argumentos considerados para disparos neuronais. Um estímulo externo $h_i$ é aplicado em um neurônio que está no seu estado inativo $x_i = 0$, com o estímulo o neurônio dispara, tornando-se ativo ($x_i = 1$). Após o disparo o neurônio permanecerá no estado refratário nos tempos seguintes.

Sendo assim, quando todas as regras colocadas anteriormente são associadas, tem-se um modelo de autômato celular composto por células neuronais. A medida que as células interagem, regidas por probabilidades, tem-se a formação de um tumor. O tumor afetará os neurônios indiretamente, por meio das células responsáveis por sua manutenção. Desta forma, na região onde o tumor se desenvolve, a atividade deixa de existir.

Na Figura 4, observa-se a evolução do autômato celular proposto, composto de células neuronais com funções de sustentação e alimentação (exemplo: células gliais) e por neurônios. A figura é uma imagem no instante de tempo *t=50* para uma rede de *100* x *100* neurônios. Em verde encontram-se os neurônios vivos, porém ociosos, em amarelo encontram-se os neurônios ativos (disparando) e em branco encontra-se a região onde os neurônios morreram. Além da interação entre células, a figura ainda mostra na região branca que os neurônios mortos são inativos, isto é, não disparam.

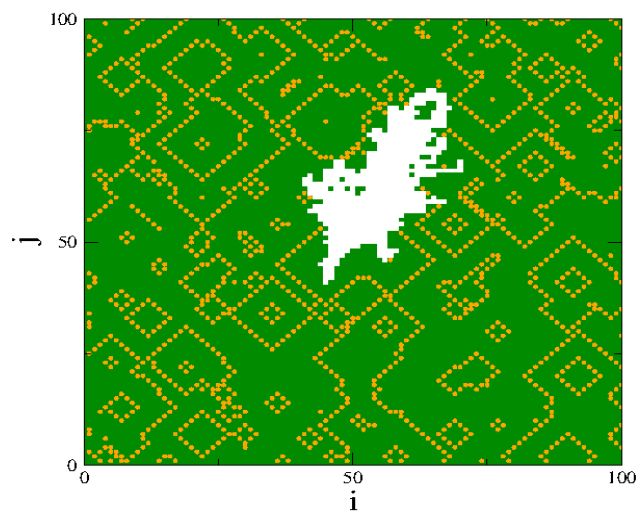

**Figura 4** – Rede neuronal com proliferação de células cancerígenas. A imagem representa o instante *t=50* de uma rede de 100 x 100 neurônios. Em verde encontram-se os neurônios vivos (ociosos), em amarelo os neurônios ativos e em branco os neurônios mortos (inativos).

### 3. Células neuronais, proliferação celular e disparos neuronais

As células apresentadas na seção 2 duplicam-se com probabilidade $p_1$, sendo as glias uma possível representação destas células. Não preocupou-se em separar as células gliais por tipos, apenas limitou-se que as células não realizam sua parte funcional para com os neurônios. Os neurônios dependem diretamente das células gliais para receber nutrientes, oxigênio, proteção contra patógenos, remoção de neurônios mortos, e para auxílio na transmissão sináptica [Laming2000, Griffith1996]. Caso as células gliais sofram algum tipo de distúrbio os neurônios são diretamente afetados.

As células gliais são células neuronais que se reproduzem, e podem sofrer distúrbios que levam proliferação não controlada [Otto2002] e podem acabar dando origem a um tumor intracranianos (o tipo de tumor depende de qual célula deu origem ao mesmo). Tais tumores podem ter origem nos diferentes tecidos cerebrais (tumores primários) ou serem provenientes de tumores malignos em outros órgãos (tumores secundários ou metastáticos). Ambos os tipos de tumores tem diferentes efeitos no organismo, porém, em geral podem causar cefaleia, vômitos, perda de coordenação motora, perda de visão e audição, perda de força e sensibilidade, alterações comportamentais, crises convulsivas e alterações nas atividades neuronais [Hahn2002, Araújo1998, Costa2000, Andrade2002, Penny2007].

Durante o processo de proliferação das células cancerígenas, o organismo tenta defender-se usando o sistema imunológico, que tentará aniquilar as células cancerígenas. Nessa tentativa existe uma probabilidade de sucesso, mas também existe a probabilidade de que algumas células sobrevivam e tornem-se resultado desse processo citotóxico. As células citotóxicas tem capacidade destrutiva por meio da liberação de substâncias nocivas [Keener2009, Souba1993, Lyons2007].

Os neurônios e as células glias funcionam de modo integrado, formando circuitos neurônio-glias que processam informações vindas de ambientes externos e internos, bem como do próprio sistema nervoso [Fieldes2006]. Tanto o neurônio quanto as glias podem gerar sinais de informação. Entretanto, somente o neurônio gera sinais bioelétricos integrados às vias de sinalização bioquímica de seu citoplasma [Lent2010]. O sistema de comunicação ao entre neurônios e células glias é bidirecional e em média existem dez células glias ligadas a um neurônio [Kandel2000]. A atividade dos neurônios disparando pode ser quantificada, para tanto realiza-se medidas da razão entre densidade média de disparos neuronais em um período de tempo [Bear2008].

**4. Parâmetros utilizados no modelo**

Para as simulações considera-se os parâmetros a partir de resultados de trabalhos experimentais sobre a dinâmica do crescimento de células ativas [Otto2002] realizados por Qi e colaboradores [Qi1993]. Na Tabela 1, encontram-se os intervalos de valores para as probabilidades que influenciam diretamente o modelo. Na Tabela também é possível verificar o valor de densidade crítica que deixa o modelo com coerência biológica e se justifica por ser o quanto de células cancerígenas está ocupando o espaço amostral limitado por uma constante φ.

**Tabela 1** – Intervalos de probabilidade que influenciam diretamente o modelo.

| Parâmetro | Intervalo |
|-----------|-----------|
| $p_1$ | 0,26 – 0,48 ou 0,58 – 0,89 |
| $p_2$ | 0,2 – 0,4 |
| $p_3$ | 0,2 – 0,65 |
| $p_4$ | 0,1 – 0,4 |
| $d_c$ | 3,85 |
| $\varphi$ | $10^3$ |

## 5. Resultados do modelo

Como pode ser notado na Figura 4, conforme um tumor desenvolve, neurônios morrem e a atividade neuronal deixa de existir no local. Nesta seção, apresenta-se o que ocorre com a evolução temporal da taxa de disparos dos neurônios dessa rede. Para as simulações foram utilizados os parâmetros da Tabela 1. Inicialmente apenas 5 células são dispostas como cancerígenas e as demais surgem de acordo com a evolução do autômato.

Na Figura 5 representa-se um caso onde as probabilidades $p_1 = p_2 = 0$, ou seja, as células cancerígenas não estão se duplicando ou sofrendo alguma transformação. Considera-se no modelo uma quantidade inicial de células cancerígenas, no caso desta simulação são 5 células cancerígenas, as quais que com estes valores dos parâmetros não irão sofrer duplicação. Pode-se dizer que essa quantidade representa um valor mínimo de células cancerígenas, o que ainda não ocasionou redução de neurônios. Aos neurônios aplica-se perturbações externas diferentes, limitando estes valores em 0,01 e 100, ou seja, limita-se a análise de mais neurônios disparando para as mesmas probabilidades de proliferação.

Na Figura 5 (a) tem-se uma perturbação externa $r = 0,01$ para qual a taxa de disparo é estável. Na figura 5 (b) o valor da perturbação da rede atinge um limiar $r = 100$ em relação a quantidade de neurônios e as probabilidades atribuídas a simulação. Nesse caso a rede comporta-se como se todos os neurônios estivessem sendo perturbados ao mesmo tempo e depois ficassem todos quiescentes.

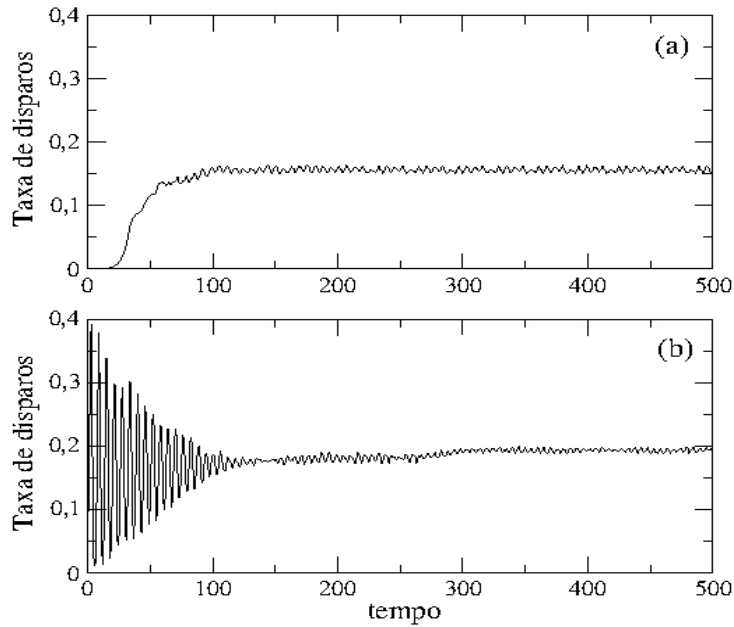

**Figura 5** – Taxa de disparos neuronais para um autômato com $N = 100$, 5 células cancerígenas evoluindo com probabilidades de proliferação $p_1 = p_2 = 0$, $p_3 = p_4 = 0,4$, densidade crítica de células cancerígenas $d_c = 3,85$ e diferentes perturbações externas, em (a) $r = 0,01$ e (b) $r = 100$.

Como apresentado anteriormente, a taxa de disparos neuronais de uma rede sem a presença de proliferação de células cancerígenas é afetada pelo valor de perturbação externa atribuída a rede neuronal. Neste momento um segundo estudo é realizado, apresenta-se neste a evolução temporal da taxa de disparos para situações considerando probabilidades de duplicação das células cancerígenas e perturbações externas em seus limiares mínimos e máximos [Qi1993, Copelli2002, Iarosz2011, Iarosz2012].

Na Figura 6 é possível notar a dependência dos disparos neuronais com relação a proliferação das células cancerígenas e a perturbação externa [Iarosz2011]. Nas Figuras 6 (a) e (b) apresenta-se perturbação externa $r = 0,01$ com probabilidades de duplicação das células cancerígenas $p_1 = 0,2$ e $p_1 = 0,68$, respectivamente. Nas Figuras 6 (c) e (d) tem-se $r = 100$, onde considerou-se probabilidades de duplicação $p_1 = 0,2$ e $p_1 = 0,68$, respectivamente. Na Figura 6 (a) a taxa de disparos aumenta de valor no início, e após este transiente a rede neural passa a apresentar pequenas oscilações próximas de 1,9. Em 6 (b) considera-se um alto valor para a probabilidade de duplicação das células cancerígenas, o que resulta em um significativo decréscimo do número de neurônios, desta forma a medida da taxa de disparos neuronais decresce com a evolução temporal.

Nos seguintes gráficos da Figura 6 encontram-se os casos com probabilidades de duplicação das células cancerígenas $p_1 = 0,2$ para (c) e $p_1 = 0,68$ para (d). Em ambos os casos a perturbação externa encontra-se no limiar máximo para a rede proposta neste autômato $r = 100$. Observa-se que independente do valor de $p_1$, valores de perturbação externa elevados, levam a rede ao comportamento coletivo. Observando a Figura 6 é conclusivo que a proliferação de células cancerígenas influência diretamente a atividade neuronal.

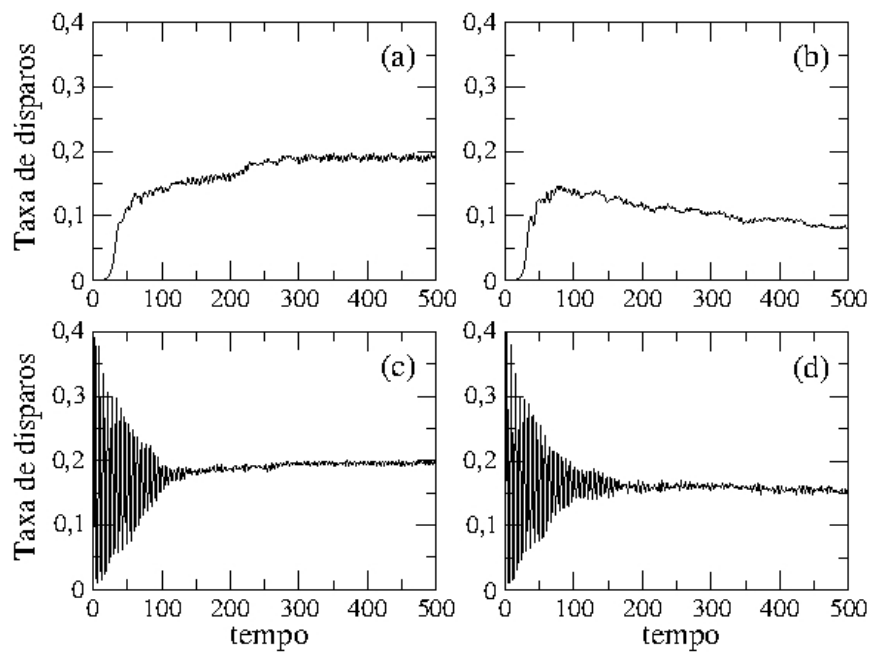

**Figura 6** – Taxa de disparos neuronais para um autômato com $N = 100$, 5 células cancerígenas evoluindo com probabilidades de proliferação $p_2 = 0$, $p_3 = p_4 = 0,4$, densidade crítica de células cancerígenas $d_c = 3,85$, perturbações externas $r = 0,01$ em (a) e (b), $r = 100$ em (c) e (d), probabilidade de duplicação das células cancerígenas $p_1 = 0,2$ para (a) e (c), e $p_1 = 0,68$ para (b) e (d).

## 6. Conclusões

Neste trabalho estudou-se como a taxa de disparos neuronais é afetada pelo crescimento do câncer. Considerou-se o modelo de autômato celular, visto que é um modelo simples e eficiente em simulações envolvendo proliferação celular e rede neuronal. Com as simulações realizadas neste trabalho observou-se que o modelo de autômato celular apresenta dependência com as condições iniciais, isto é, a probabilidade de proliferação de células cancerígenas influencia a taxa de disparos neuronais, bem como valores extremos de perturbações externas causam efeitos coletivos na rede, que no entanto, estabilizam para um valor assintótico com o passar do tempo.

Quanto maior a perturbação externa observa-se uma maior taxa de disparos. Verificou-se alterações na estatística da taxa de disparos devido ao crescimento do câncer. Quanto maior o câncer, menor a taxa de disparos neuronais.

**Agradecimentos**

**7. Referências**